\documentclass[%
reprint,
unsortedaddress,
bibnotes,
amsmath,amssymb,
aps,
 floatfix,
twocolumn
]{revtex4}

\usepackage{graphicx}
\usepackage[space]{grffile}
\usepackage{latexsym}
\usepackage{amsfonts,amsmath,amssymb}
\usepackage{url}
\usepackage[utf8]{inputenc}
\usepackage{hyperref}
\hypersetup{colorlinks=false,pdfborder={0 0 0}}
\usepackage{textcomp}
\usepackage{longtable}
\usepackage{multirow,booktabs}

\begin{document}

\title{
Quantum Monte Carlo with very large multideterminant wavefunctions
}

\author{Anthony Scemama$^1$ and Thomas Applencourt$^1$ and Emmanuel Giner$^2$ and Michel Caffarel$^1$}
\affiliation{$^1$Lab. Chimie et Physique Quantiques, CNRS-Universit\'e de Toulouse, France.}
\affiliation{$^2$Dipartimento di Scienze Chimiche e Farmaceutiche, Università degli Studi di Ferrara, Ferrara, Italie.}

\begin{abstract}

An algorithm to compute efficiently the first two derivatives of (very) large
multideterminant wavefunctions for quantum Monte Carlo calculations is presented.
The calculation of determinants and their derivatives is performed using 
the Sherman-Morrison formula for updating the inverse Slater matrix. 
An improved implementation based on the reduction 
of the number of column substitutions and on a very efficient implementation of 
the calculation of the scalar products involved is presented.
It is emphasized that multideterminant expansions contain in general
a large number of identical spin-specific determinants: for 
typical configuration interaction-type wavefunctions
the number of unique spin-specific determinants $N_{\rm det}^\sigma$ ($\sigma=\uparrow,\downarrow$) 
with a non-negligible weight in the expansion is of order ${\cal O}(\sqrt{N_{\rm det}})$.
We show that a careful implementation of the calculation of the $N_{\rm det}$-dependent 
contributions can make this step negligible enough 
so that in practice the algorithm scales as the total 
number of unique spin-specific determinants,$\; N_{\rm det}^\uparrow + N_{\rm det}^\downarrow$,
over a wide range of total number of determinants (here, $N_{\rm det}$ 
up to about one million), thus greatly reducing the total computational cost.
Finally, a new truncation scheme for the multideterminant expansion is proposed so that
larger expansions can be considered without increasing the computational time.
The algorithm is illustrated with all-electron
Fixed-Node Diffusion Monte Carlo calculations of the total energy of the chlorine atom.
Calculations using a trial wavefunction including about 750~000
determinants with a computational increase of $\sim$ 400 compared
to a single-determinant calculation are shown to be feasible.
\end{abstract}

\maketitle

\section{Introduction}
\label{intro}

In a series of recent works we have proposed to use very large 
configuration interaction (CI) trial wave functions in Fixed-Node 
Diffusion Monte Carlo (FN-DMC).\cite{Giner_2015,Scemama_2014,Giner_2013,Caffarel_2014} 
The main bottleneck of such calculations is the price to pay for computing 
the first two derivatives of the trial wavefunction at each Monte Carlo step. 
In the present paper, we describe in detail the various strategies we have 
devised to make such calculations feasible. To illustrate quantitatively 
the performance of our algorithm,
let us mention that in a first application to the oxygen atom,\cite{Giner_2013} 
converged all-electron Fixed-Node DMC calculations have been possible with a 
trial wavefunction including up to 100~000 Slater determinants.
In another application to the metal atoms of the $3d$ series,\cite{Giner_2015} 
up to about 48 000 determinants for all-electron FN-DMC simulations have been used.
In the illustrating case of the chlorine atom used here, 
a converged all-electron Fixed-Node 
DMC calculation including up to 750~000 Slater determinants is presented.
The trial wavefunction and its derivatives being expressed as a sum of determinants, 
the computational time needed at each Monte Carlo step is expected to scale linearly in the number 
of determinants, a situation which can rapidly become intractable if large expansions are 
desired (say, greater than a few thousands). 

To tackle this difficulty a number of methods have been recently proposed. Nukala and Kent
\cite{Nukala_2009} have introduced a recursive algorithm for updating the Slater determinants 
reducing significantly the computational complexity.
In 2011, Clark {\em et al.} have proposed the Table method\cite{Clark_2011} which leads 
to a total cost per step scaling as 
${\cal O}(N_{\rm elec}^2) + {\cal O}(N_s N_{\rm elec}) + {\cal O}(N_{\rm det})$, $N_{\rm elec}$ being the number of electrons, $N_{s}$ the number 
of single excitations, and $N_{\rm det}$, the number of determinants. 
In the case of the water molecule 
a speedup of about 50 has been obtained. 
More recently, Weerasinghe {\em et al.} proposed a compression
algorithm\cite{Weerasinghe_2014} based on the idea of reducing the number of determinants
of the expansion by combining repeatedly determinants differing by one single orbital.
The total cost per step is then reduced by about the compression factor, that is the ratio of
the initial to the final number of determinants. When applied to the first-row atoms,
a reduction of the number of determinants by a factor of up to about 27 has been obtained (however,
for the sake of comparison this factor should be reduced here to about 11, see note \cite{note1}).

Here, we present our approach to reduce the computational cost of large multideterminant expansions, 
as it is implemented in the QMC=Chem program developed in our group.\cite{Scemama_2013} 
Before describing our algorithm it is important to emphasize on a fundamental and 
general aspect regarding efficient calculations of large multi-determinant expansions. 
Chemical systems studied in quantum chemistry are in general compact 
(not extended over large portion of space like, for example, 
in solid-state applications) and include a fixed and moderate number 
of electrons (say, up to a thousand of active electrons). 
It is thus important to be very cautious with the notion of 
scaling law of the computational effort as a function of the various critical parameters: Number 
of electrons, orbitals, and determinants. Indeed, the 
asymptotic regime where such scaling laws are valid is not necessarily reached in practice and 
the prefactors generally play a crucial role. In particular, the algorithm leading to the best 
{\it theoretical} scaling law is not necessarily {\it in practice}
the most optimal one, as will be the case here.
Following this idea, our algorithm has been designed to be a good compromise between good 
scaling laws and efficient practical implementation on modern processors. 
A detailed discussion of this important aspect is presented in section\ref{sec5}.

The standard approach in QMC is to use a Slater-Jastrow trial wavefunction consisting of 
a determinantal CI-type expansion and a Jastrow prefactor to describe the explicit electron-electron 
and electron-electron-nucleus interactions (dynamical correlation effects). The Jastrow factor 
can be treated individually using standard techniques 
with a computational cost of about ${\cal O}(N_{\rm elec}^2)$
or even less if unphysical long-range interactions are cut off. In what follows, the Jastrow 
prefactor will thus be omitted for simplicity.

A general $N$-electron and $N_{\rm det}$-determinant CI wave function $\Psi$ can be expressed 
in the spin-free formalism used in QMC, denoting  ${\bf R}=({\bf r}_1,...,{\bf r}_N)$ the full set of
electron space coordinates, and ${\bf R}_\uparrow$ and ${\bf R}_\uparrow$ the
two subsets of coordinates associated with $\uparrow$ and $\downarrow$ electrons~:
\begin{equation}
\Psi({\bf R})=\sum_{k=1}^{N_{\rm det}} c_k 
{\rm Det} {\bf S}^\uparrow_k({\bf R}_{\uparrow}) {\rm Det} {\bf S}^\downarrow_k ({\bf R}_{N_\downarrow})
\label{eq1}
\end{equation}
In this formula, the matrix 
elements of the $N_\sigma \times N_\sigma$ Slater matrices ${\bf S}^\sigma_k$ ($\sigma=\uparrow,\downarrow$) 
are defined as $[{\bf S}^\sigma_k]_{ij}=\phi_j({\bf r}_i)$ where $\phi_j$ are single-particle molecular orbitals 
(for simplicity, a common set of orbitals for $\uparrow$- and $\downarrow$- electrons is used here, 
the generalization to two different sets being straightforward).
The Slater matrices are labelled by an integer $k$ defining which 
specific subset of $N_\sigma$ molecular orbitals is used to build it, the molecular orbitals 
being chosen among a set of $N_{\rm MO}$ active orbitals [$\binom {N_{\rm MO}}{N_\sigma}$ such 
possibilities]. It must be emphasized that expansion (\ref{eq1}) contains in general a large number 
of identical spin-specific determinants (in the case of the chlorine atom treated here,
only 35 584 determinants are different out of the 1 000 000 determinants of the expansion, 
see Table~\ref{tab:cl_dz:2}). 
In practice, it is then important to calculate only once these unique determinants.
A convenient form for the CI wavefunction taking account of this aspect
is the following bilinear form
\begin{equation} \label{eq:main}
\Psi({\bf R})  = \sum_{i=1}^{N_{\rm det}^\uparrow} \sum_{j=1}^{N_{\rm det}^\downarrow} C_{ij} 
D^\uparrow_i({\bf R}_\uparrow) D^\downarrow_j({\bf R}_\downarrow)  
= {{\bf D}_\uparrow}^\dagger {\bf C} {\bf D}_\downarrow.
\end{equation}
where ${{\bf D}_\sigma }$ are column vectors containing the values at the $\sigma-$electron
positions of the $N_{\rm det}^\sigma$ {\it different} 
determinants appearing in the $N_{\rm det}-$expansion, Eq.(\ref{eq1}), and ${\bf C}$ is the matrix of 
coefficients of size $N_{\rm det}^\uparrow \times N_{\rm det}^\downarrow$.

In practical applications the wavefunction (\ref{eq:main}) is rarely 
the full CI expansion (except for small systems or very small basis set) but some 
approximate form resulting usually from the truncation of a limited configuration 
interaction (typically, CISD) or CASSCF calculation using a threshold 
for determinantal coefficients. In the case of the Full Configuration Interaction (FCI) 
where all possible determinants are considered,
 $N_{\rm det}^\sigma$ attains its maximal value of $\binom {N_{\rm MO}}{N_\sigma}$.
In that case, $N_{\rm det} = N_{\rm det}^\uparrow \times N_{\rm det}^\downarrow$ and 
the number of unique spin-specific determinants $D^\sigma({\bf R})$ is 
of order $\sqrt{N_{\rm det}}$. In practice, using truncated forms 
the number of unique determinants of non-negligible weight also follows a 
similar rule, essentially because the most numerous excitations implying multiple 
excitations of electrons of same spin plays physically a marginal role 
(see note \cite{notesqrt} for a more quantitative discussion).
As a consequence, we note 
that the constant coefficient matrix ${\bf C}$ is usually (very) sparse since it contains
$N_{\rm det}$ non-zero entries where $N_{\rm det} \ll N_{\rm det}^\uparrow \times
N_{\rm det}^\downarrow$. 

In section~\ref{sec2} all theoretical and practical details of our algorithm is presented. 
In section~\ref{sec3}, a new truncation scheme for the 
CI expansion motivated by the structure of the bilinear form is proposed. This truncation scheme 
allows to compute less $\uparrow$- and $\downarrow$- determinants  
than in the standard procedure where 
a threshold is applied on the coefficients $|c_k|$ of Eq.(\ref{eq1})
(see, {\it e.g.} \cite{Clay_2015}).
In section~\ref{sec4}, numerical results for the chlorine atom are presented to illustrate
the various aspects of the algorithm using CI wave functions containing up to one million of Slater
determinants. In this example it is shown that 
the FN-DMC energy can be obtained with a trial wavefunction including about 750~000
determinants with a computational increase of only $\sim$400 compared
to the same calculation using a single determinant.
In section~\ref{sec5} a number of important remarks we believe to be important to take into account 
when comparing and implementing different algorithms are made. 
As an illustration, a comparison of the performance of 
our algorithm compared to that of the Table method of Clark {\it et al.},\cite{Clark_2011} 
is presented.
Finally, a summary of our main results is presented in Sec.6.

\section{Algorithm and implementation}
\label{sec2}
At every Monte Carlo step, the values of all the $N_{\rm MO}$ 
molecular orbitals (MOs) are computed (or a subset if some orbitals are never used in $\Psi$)
at all electron positions and stored
in an $(N_{\rm elec}^\uparrow+N_{\rm elec}^\downarrow)\times N_{\rm MO}$ array $\Phi$.
Similarly, the derivatives of the MOs with respect to the electron coordinates
(gradients $\nabla_{x,i}$, $\nabla_{y,i}$, $\nabla_{z,i}$, and Laplacian
$\Delta_{i}$) are stored in four arrays $\nabla_x \Phi$, $\nabla_y \Phi$,
$\nabla_z \Phi$, and $\Delta \Phi$. Our implementation has already been
detailed in reference~\citep{Scemama_2013_2}, but let us recall that this step
can be computed very efficiently on modern x86 central processing units (CPUs)
as it makes an intensive use of vector fused multiply-add (FMA) instructions
and has a very low memory footprint.
In this section, we describe how the multi-determinant wave function is
evaluated, as well as its derivatives (gradients and Laplacian).

\subsection{Pre-processing}
\label{subsec:Pre-processing}
To take full advantage of the bilinear form, Eq.(\ref{eq:main}), a preliminary step to be done only
once before the QMC run is performed. The purpose is twofold.
First, to define a convenient encoding of the determinants making their manipulation easy and 
very rapid, and their storage requirements very low.
Second, to introduce a comparative function allowing to sort the determinants 
so that contiguous determinants in the sorted list are likely to have a 
small number of differences in terms of multiple-particle excitations. This step will be important 
to minimize the number of Sherman-Morrison updates of the Slater matrices as discussed in the 
next section.

{\it Encoding}. Determinants are initially encoded using 64-bit integers as described in
reference~\citep{slater_condon}: When the number of MOs is less or equal to 64, one
integer encodes the occupation of the orbitals by the $\uparrow$ electrons and another
one encodes the occupation of the orbitals by the $\downarrow$ electrons,  by setting to one
the bits corresponding to the positions of occupied MOs.
For instance, the Hartree-Fock determinant for the chlorine atom (9
$\uparrow$-electrons and 8 $\downarrow$-electrons) is encoded as (511,255), which
is in binary representation ({\tt 111111111},{\tt 11111111}),
and the doubly-excited determinant resulting from
an excitation from the MO \#7 to \#12 for a $\uparrow$ and $\downarrow$ electron 
is (2495,2239) or
({\tt 100110111111},{\tt 100010111111}).
When the system contains more than 64 MOs, several $N_{\rm int}$ 64-bit
integers are used for each spin-specific determinant. The initial storage
requirement is therefore 
$N_{\rm det} \times (\lfloor N_{\rm MOs}/64 \rfloor + 1) \times 16$ bytes. 

The $\uparrow$ and $\downarrow$ determinants are treated in two distinct lists.
Each spin-specific list is then treated independently as follows.

{\it Sorting of determinants} The list of determinants is sorted with respect to some comparison function. 
We recall that in a sort algorithm a key is associated with each element of the list and that 
the choice of the comparison function is not unique. Furthermore, an exact mapping between 
the elements of the list and the values of the key is not necessary (several determinants 
can have a common key). We have tested a variety of keys with the objective of having both a 
simple and efficient encoding and an ordered list of determinants where contiguous
determinants have a minimal number of differences in terms of particle-excitations
with high probability.

\begin{table}
\begin{tabular}{|c|c|c|c|}
\hline
List index & Decimal  &     Binary     &  Determinant    \\ 
\hline
     1      &    15    & {\tt 00001111} & $|1234\rangle$  \\ 
     2      &    23    & {\tt 00010111} & $|1235\rangle$  \\ 
     3      &    27    & {\tt 00011011} & $|1245\rangle$  \\ 
     4      &    29    & {\tt 00011101} & $|1345\rangle$  \\ 
     5      &    30    & {\tt 00011110} & $|2345\rangle$  \\ 
     6      &    39    & {\tt 00100111} & $|1236\rangle$  \\ 
     7      &    43    & {\tt 00101011} & $|1246\rangle$  \\ 
     8      &    45    & {\tt 00101101} & $|1346\rangle$  \\ 
  $\cdots$  & $\cdots$ &    $\cdots$    &   $\cdots$      \\
\hline
\end{tabular}
\caption{\label{tab:ordering}Ordering of determinants given by Eq.~(\ref{eq:comp}).}
\end{table}

The key $\omega$ used here
is the numerical value of the 64-bit integer obtained by accumulating an {\tt
xor} operation ($\oplus$) on all the $N_{\rm int}$ 64-bit integers $i_n$
constituting the determinant
\begin{equation}
  \label{eq:comp}
  \omega = i_1 \oplus \dots \oplus i_{N_{\rm int}} - 2^{63}
\end{equation}
As Fortran does not handle unsigned integers, we shift the value by $-2^{63}$
to get an ordering consistent with the unsigned representation.
Table~\ref{tab:ordering} gives an example of the ordering with 4 electrons in 8
orbitals. One can remark that the probability of using a single excitation to
go from one determinant to the next one in the list is very high.
The sort is performed in a linear time with respect to $N_{\rm
det}^\uparrow$ and $N_{\rm det}^\downarrow$ thanks to the radix sort
algorithm.\cite{Andersson_1995}
Then, duplicate determinants are filtered out by searching for duplicates among
determinants giving the same key $\omega$. At this point, we have two
spin-specific lists of sorted determinants containing respectively $N_{\rm
det}^\uparrow$ and $N_{\rm det}^\downarrow$ unique determinants.

{\it Sparse representation of $C_{ij}$}. We now want to express the matrix of coefficients ${\bf C}$ 
in a sparse coordinate format made of an 
array of values, an array of column indices, and an array of row indices.
Note that the dimension of such arrays is exactly $N_{\rm det}$.
For each determinant product in Eq.\ref{eq:main}, we compute the key $\omega$ corresponding to the
$\uparrow$ determinant. As the list of unique determinants is sorted, we can use
a binary search to find its position $i$ in the list in logarithmic time. This
position is appended to the list of row indices. Similarly, the list of
column indices is updated by finding the position $j$ of the $\downarrow$
determinant. To improve the memory access patterns in the next steps, the value
$N_{\rm det}^\uparrow \times (j-1)+i$ is appended to an additional temporary
array.
Finally, the additional temporary array is sorted (in linear time with the
radix sort), and we apply the corresponding ordering to the three arrays
containing the sparse representation of the ${\bf C}$ matrix. Now, the elements
of the ${\bf C}$ matrix are ordered such that reading the arrays sequentially
corresponds to reading the matrix column by column.

Let us emphasize that this pre-processing step is not a bottleneck 
as it scales linearly with the number of determinant products and has to 
be done only once. For instance,
this pre-processing step takes roughly 3 seconds on a single core for a 
wave function with one million of Slater determinants.
In sharp contrast, the computations described in the next paragraphs that need to be
performed at every Monte Carlo step are critical.

\subsection{Calculation of the vectors ${\bf D}_\uparrow$ and ${\bf D}_\downarrow$}
\label{calc}
The list of integers corresponding to the indices of the molecular orbitals
occupied in the first determinant $D^\uparrow_1$ is decoded from its compressed
64-bit integer representation.
This list is used to build the Slater matrix ${\bf S}^\uparrow_1$ corresponding
to $D^\uparrow_1$ by copying the appropriate $N_{\uparrow}$ columns of $\Phi$.
We then evaluate the determinant and the inverse Slater matrix in the usual
way: we perform the LU factorization of ${\bf S}^\uparrow_1$ using partial
pivoting (using the {\tt dgetrf} \textsc{lapack} routine\cite{lapack}, 
${\cal O}({N_{\rm elec}^\uparrow}^3)$). It is now
straightforward to obtain the determinant $D^\uparrow_1({\bf R}_\uparrow)$,
and the inverse Slater matrix $\left({\bf S}^\uparrow_1 \right)^{-1}$ is
obtained using the {\tt dgetri} \textsc{lapack} routine in
${\cal O}({N_{\rm elec}^\uparrow}^3)$.
If the dimension of the Slater matrix is smaller than $6 \times 6$, one can
remark that this cubic algorithm will cost more than the
na\"ive ${\cal O}(N_{\rm elec}^\uparrow!)$ algorithm. Moreover, linear algebra
packages are
optimized for large matrices and usually do not perform well on such small
matrices. Therefore, we used a script to generate hard-coded subroutines
implementing the na\"ive algorithm for the calculation of the determinant and
the inversion of $1\times 1$ to $5\times 5$ matrices.

For all the remaining determinants $\{D_{i>1}^\uparrow \}$, the Sherman-Morrison (SM)
formula is used to update the inverse Slater matrix {\em in place} in ${\cal
O}({N_{\rm elec}^\uparrow}^2)$. 
The column updates are executed sequentially by substituting
one column at a time. In the case of a double excitation for instance, a
sequence of two updates will be performed. The substitution taking place at $k$-th column, 
the SM formula is given by
\begin{equation}
  [({\bf S}+{\bf u} {\bf v}_k^\dagger)^{-1} =
    {\bf S}^{-1} - {{\bf S}^{-1} {\bf u} {\bf v}_k^\dagger {\bf S}^{-1} \over 
    1 + {\bf v}_k^\dagger {\bf S}^{-1}{\bf u}}
\label{sherman}
\end{equation}
where ${\bf u}$ is the column vector associated with the substitution of molecular orbital $j$
by molecular orbital $j^\prime$, $u_i=\phi_{j^\prime}({\bf r}_i)- \phi_j({\bf r}_i)$
and ${\bf v}_k^\dagger= (0,\dots,1,\dots,0)$, the value 1 being at position $k$.
Other implementations\cite{Clark_2011} compare
the Slater matrix to a common reference, but here we perform the
SM updates with respect to the previously computed determinant
$D_{i-1}^\uparrow$.  
To avoid the propagation of numerical errors, we do  the following for each
$D_i^\uparrow$. If the absolute value of the ratio of the determinant with
the substituted column over the previous determinant is below $10^{-3}$, the
current column substitution is not realized and stored in a list of failed
updates.  When all updates have been tried, the list of updates to do is
overwritten by the list of failed updates and all the remaining updates are
tried again, until the list of failed updates becomes empty. If at one
iteration the length of the list of failed updates has the same non-zero length
as in the previous iteration of the sequence, the SM updates are
cancelled and the determinant is re-computed with the ${\cal O}({N_{\rm
elec}^\uparrow}^3)$ algorithm.

The SM updates are hot spots in large multi-determinant
calculations, so some particular effort was invested in their computational
efficiency.
One can first remark that it is more efficient to use the hard-coded na\"ive
algorithm to compute fully the inverse matrix from scratch than to do the
SM update for Slater matrices with dimensions $2 \times 2$ and $3
\times 3$ (the cost of the SM update is quadratic with the size of
the matrix).
Therefore, the SM updates are used for sizes greater than $3 \times 3$.
Secondly, if $N_{\rm elec}^\uparrow$ is small (typically less than 50), a general routine
is very likely to be inefficient: for example, in double loops over $i$ and $j$ running 
from 1 to {\tt n}
the compiler is not aware of the number of loop cycles {\tt n} at compile time,
so it will generate code to try to vectorize the loops (peeling loop, scalar
loop, vector loop and tail loop) and test which branch to choose at execution
time.  If the loop count is low,
the overhead dramatically affects the performance. For all matrix sizes in the
$[ 4 \times 4 : 50 \times 50 ]$ range, we have generated size-specific subroutines
from a template were the loop counts and matrix dimensions are hard-coded, 
in such a way to force the compiler to generate 100\% vectorized loops.
When needed, the tail loops are written explicitly.
The binary code produced by the compiler was validated
with the \textsc{maqao}\cite{maqao} static analysis tool by checking that vector
fused-multiply-add (FMA) instructions were produced extensively in the
innermost loops of the Sherman-Morrison updates.
For larger matrix sizes, a general subroutine is used.
In all the different versions, we use padding in the matrices to enforce the
proper memory alignment of all the columns of the matrices to enable
the vectorization of the inner-most loops without the peeling loop.

The scaling of this step is ${\cal O}({N_{\rm elec}^\uparrow}^2 \times N_{\rm
det}^\uparrow)$.

\subsection{Calculation of the gradients and Laplacian}

The bilinear expression of the wave function in Eq.(\ref{eq:main}) yields
the following expressions for the derivatives:

\begin{eqnarray}
 \label{eq:grad_psi}
  \nabla_{x,i} \Psi & = & (\nabla_{x,i} {{\bf D}_\uparrow})^\dagger {\bf C} {\bf D}_\downarrow + {{\bf D}_\uparrow}^\dagger {\bf C} (\nabla_{x,i} {\bf D}_\downarrow ) \\
 \label{eq:lapl_psi}
 \Delta_i \Psi & = & (\Delta_i {{\bf D}_\uparrow})^\dagger {\bf C} {\bf D}_\downarrow + {{\bf D}_\uparrow}^\dagger {\bf C} (\Delta_i {\bf D}_\downarrow ) + 2 [ \\ 
               && ({\nabla_{x,i} {\bf D}_\uparrow})^\dagger {\bf C} (\nabla_{x,i} {\bf D}_\downarrow) + \nonumber \\
               && ({\nabla_{y,i} {\bf D}_\uparrow})^\dagger {\bf C} (\nabla_{y,i} {\bf D}_\downarrow) + \nonumber \\
               && ({\nabla_{z,i} {\bf D}_\uparrow})^\dagger {\bf C} (\nabla_{z,i} {\bf D}_\downarrow) ] \nonumber 
\end{eqnarray}

In the expression of the Laplacian of the wave function
(Eq.(\ref{eq:lapl_psi})), the gradient terms $\nabla_{i}{\bf D}_\uparrow$ vanish
when $i$ is a $\downarrow$ electron. Similarly, the terms $\nabla_{i}{\bf D}_\downarrow$
vanish when $i$ is an $\uparrow$ electron.  As a consequence, the cross-terms
involving
both the gradients $\nabla_{i}{\bf D}_\uparrow$ and $\nabla_{i}{\bf D}_\downarrow$
are always zero, and the $3N_{\rm elec}$ components of the gradient and
Laplacian can be computed using the same instructions with different
input data. Hence we define ${\tilde \nabla}_i$ as a four-element vector
$[ \nabla_{x}, \nabla_{y}, \nabla_{z}, \Delta ]$, and one only needs to implement

\begin{equation}
{\tilde \nabla}_i \Psi = ({\tilde \nabla}_i {{\bf D}_\uparrow})^\dagger {\bf C} {\bf D}_\downarrow + {{\bf D}_\uparrow}^\dagger {\bf C} ({\tilde \nabla}_i {\bf D}_\downarrow )
\label{eq:nabla}
\end{equation}

The gradients and the Laplacian of the wave function are computed together using
Eq.(\ref{eq:nabla}) in an array of dimension $4 \times N_{\rm elec}$, using
the arrays ${\tilde \nabla} {\bf D}^\uparrow$ and ${\tilde \nabla} {\bf
D}^\downarrow$, dimensioned respectively as $4 \times N_{\rm elec}^\uparrow \times
N_{\rm det}^\uparrow$ and $4 \times N_{\rm elec}^\downarrow \times N_{\rm det}^\downarrow$.
The computation of all the four components of ${\tilde \nabla} \Psi$ can be performed
simultaneously using Single Instruction Multiple Data (SIMD) vector
instructions.  Indeed, modern x86 CPUs can use vector operations on 256- or 512
bit-wide vectors, which correspond to 4 or 8 double precision elements, if the
arrays are properly aligned in memory. Hence we aligned the arrays on 512-bit
boundaries using compiler directives.

The ${\tilde \nabla} {\bf D}_j^\uparrow$ are computed using the array ${\tilde
\nabla}\Phi$ and the inverse Slater matrix:
\begin{equation}
  \label{eq:grap_psi_psi_inv}
  {\tilde \nabla}_i {\bf D}_j^\uparrow = \sum_k {[{{\bf S}^{\uparrow}}^{-1}]_{ik}} {\tilde \nabla} {\Phi_{kj}^\uparrow}
\end{equation}

The innermost loop is the loop over the 4 components (gradients and Laplacian)
of ${\tilde \nabla}$, so we unroll twice the loop over $k$ to enable vector
instructions also on the AVX-512 micro-architecture which requires 8 double
precision elements.

As in the case of the calculation of the determinants, the scaling of this step 
is ${\cal O}({N_{\rm elec}^\uparrow}^2 \times N_{\rm
det}^\uparrow)$.

The calculation of the derivatives of the total wave function is then performed using two
dense matrix-vector products: 
$(\nabla {{\bf D}_\uparrow})^\dagger \cdot ({\bf C} {\bf D}_\downarrow)$ and
$({{\bf D}_\uparrow}^\dagger {\bf C}) \cdot (\nabla {\bf D}_\downarrow )$, as detailed
in the next subsection.

\subsection{Computation of the intermediate vectors and $\Psi$}
\label{subsec:comput-intermediate-vectors}
An important point is that the two matrix-vector products ${{\bf D}_\uparrow}^\dagger
{\bf C}$ and ${\bf C} {\bf D}_\downarrow$ need to be performed only once, and the
resulting vectors are used for the computation of $\Psi$ and ${\tilde \nabla}\Psi$.
As this step consists in two sparse matrix/dense vector product, it has inevitably a low
arithmetic intensity (small number of floating point operations per data loaded or stored)
and the execution speed is limited by data access. It is therefore critical to optimize for this step the
data movement from the main memory to the CPU cores.
As the same matrix ${\bf C}$ is used in both products, the two products can be
computed simultaneously:
\begin{verbatim}
do k=1,det_num
  i = C_rows(k)
  j = C_columns(k)
  Da_C(j) = Da_C(j) + C(k)*Da(i)
  C_Db(i) = C_Db(i) + C(k)*Db(j)
enddo
\end{verbatim}
In this way, the three arrays corresponding to the ${\bf C}$ matrix are
streamed from the main memory through the CPU registers only once. The data 
relative to the matrix ${\bf C}$ can be moved from the main memory with
a very low latency as the hardware prefetchers of x86 CPUs are very efficient
on unit stride access patterns. Also, the ordering of the arrays of the ${\bf C}$
matrix in the pre-processing phase (see subsection about pre-processing) 
maximizes the probability of {\tt Da\_C(j)} and
{\tt Db(j)} to be in already in the CPU registers as the column index {\tt j}
is very likely to be constant from one iteration to the next. {\tt Da(i)}
and {\tt C\_Db(i)} are likely to be in a low-level cache (L1 or L2) as the
arrays are always small, dimensioned by $N_{\rm det}^\uparrow$ and $N_{\rm
det}^\downarrow$ (typically 25~KiB for a wave function with a million of Slater
determinants).

The asymptotic computational cost of this step is ${\cal O}(N_{\rm det})$. It is the 
only place in our approach where the cost is proportional to the full number of determinants.
However, thanks to the implementation just presented, the 
prefactor is so small that we have never observed that it is a time-limiting
step : in the regime where $\Psi$ has one million of determinant products,
this step takes only 10\% of the total computational time.

We chose the convention that the number of $\uparrow$ electrons is greater or
equal to the number of $\downarrow$ electrons. As a consequence the general case is
that $N_{\rm det}^\uparrow \ge N_{\rm det}^\downarrow$ so we choose to compute the
value of $\Psi$ using the dot product $({\bf D}_\uparrow {\bf C}) \cdot {\bf
D}_\downarrow$ as it involves only $N_{\rm det}^\downarrow$ operations.

\section{Improved truncation scheme}
\label{sec3}
In practice, to avoid to handle too many products of determinants in the CI 
wavefunction, Eq.(\ref{eq1}), some sort of truncation scheme is to be introduced.
In standard QMC implementations, it is usually done either by introducing 
a threshold parameter for the absolute value of the coefficients $c_k$ or
by taking the smallest number of products of determinants
contributing to a given percentage of the norm of the wavefunction. 
Note that truncating coefficients of Configuration State Functions (CSFs) 
can also be considered as an improvement as it
does not break the property of the wave function to be an eigenstate of $S^2$.

In the preceding section, it has been shown how to compute as 
efficiently as possible the derivatives of the trial wavefunction for a given 
number of products of spin-specific determinants, $N_{\rm det}$. A remarkable result is that the 
bulk of the computational effort may be reduced to the calculation of 
$\uparrow$- and $\downarrow$-determinants.
Accordingly, to remove a product of determinants 
whose spin-specific determinants are already present in other products will not 
change the computational cost. A natural idea is thus to truncate the 
wavefunction by removing {\it independently} $\uparrow$- and $\downarrow$- determinants.

To do this, we decompose the norm of the
wave function as
\begin{equation}
  {\cal N} = \sum_{i=1}^{N^\uparrow_{\rm det}} \sum_{j=1}^{N^\downarrow_{\rm det}} C_{ij}^2 = 
  \sum_{i=1}^{N^\uparrow_{\rm det}} {\cal N}_i^\uparrow = \sum_{j=1}^{N^\downarrow_{\rm det}} {\cal N}_j^\downarrow
\end{equation}
where 
${\cal N}_i^\uparrow = \sum_{j=1}^{N^\downarrow_{\rm det}} C_{ij}^2$ and ${\cal N}_j^\downarrow = \sum_{i=1}^{N^\uparrow_{\rm det}} C_{ij}^2$ are the
contributions to the norm of determinants $D_i^\uparrow$ and $D_i^\downarrow$.
We approximate the wave function by removing spin-specific determinants whose
contribution to the norm are less than a $\sigma$-dependent threshold $\epsilon_\sigma$ chosen by the user
(remove $D_k^\sigma$ such ${\cal N}_k^\sigma \le \epsilon_\sigma$).

As we shall see in the next section, this alternative truncated scheme allows 
to keep more determinants in the CI expansion at the same computational cost.

\section{Results}
\label{sec4}

The chlorine atom (17 electrons) was chosen as a benchmark. The cc-pVDZ and
cc-pVTZ basis sets\cite{Woon_1993} expressed in Cartesian coordinates
(respectively 19 and 39 molecular orbitals) have been used.
Timings were measured as the total CPU time needed for one walker to realize one
Monte Carlo step (all electrons are moved). 
It includes the calculation of the wave function, the drift
vector and the local energy.
The benchmarks were run on a single-socket desktop computer, with an
Intel\textsuperscript{\textregistered} Xeon\textsuperscript{\textregistered}
E3-1271 v3 quad-core processor at 3.60~GHz with the Turbo feature disabled.
QMC=Chem was compiled with the Intel\textsuperscript{\textregistered} Fortran
Compiler version 15.0.2 with options to generate code optimized for the AVX2
micro-architecture, and linked with the Intel\textsuperscript{\textregistered}
Math Kernel Library (MKL).
The calculation of the FN-DMC energies were performed using 800 cores on the
Curie machine (TGCC/CEA/Genci). The total computational time we used to
generate Figure~\ref{fig:energy} was 182~500 CPU hours.

Perturbatively selected configuration interaction 
wave functions of CIPSI type (Configuration Interaction using a Perturbative Selection done 
Iteratively, see ref.\cite{Giner_2013}) in the Full-CI space were
prepared with our code (Quantum Package\cite{quantum_package}) from one to one
million determinants. 

\begin{table}
\begin{tabular}{|c|c|c|c|c|c|}
\hline
$N_{\rm det}$ & $N_{\rm det}^\uparrow$ & $N_{\rm det}^\downarrow$ & $N_{\rm MOs}$ & RAM (MiB) & CPU time (ms)  \\ 
\hline
       1      &          1           &          1          &       9       &   6.12   &    0.0179    \\ 
       10      &          7           &          7          &       16      &   6.20   &    0.0470    \\ 
      100      &          40          &          32         &       18      &   6.24   &    0.1765    \\ 
     1 000     &         250          &         186         &       19      &   6.42   &    0.9932    \\ 
     10 000    &        1 143         &         748         &       19      &   7.35   &    4.5962    \\ 
    100 000    &        5 441         &        3 756        &       19      &  13.20   &   20.5972    \\ 
   1 000 000   &       21 068         &       14 516        &       19      &  45.84   &   83.1611   \\
\hline
\end{tabular}
\caption{Number of determinants ($N_{\rm det}$, $N_{\rm det}^\uparrow$ and
$N_{\rm det}^\downarrow$), number of occupied molecular orbitals ($N_{\rm MOs}$),
amount of memory per core and CPU time per per core per Monte Carlo step
({\em without} Sherman-Morrison updates). The cc-pVDZ basis set is used.}
\label{tab:cl_dz:1}
\end{table}

\begin{figure}
\begin{center}
\includegraphics[width=\columnwidth]{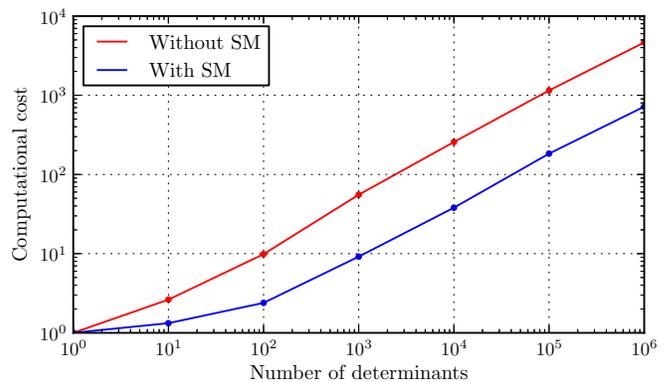}
\end{center}
\caption{
\label{fig:scaling}
Computational cost with respect to the number of determinants, normalized to
the cost of a single-determinant calculation. Results are given with and without
Sherman-Morrison updates.}
\end{figure}

To illustrate the fact that the bilinear representation of Eq.(\ref{eq:main})
gives rise to an ${\cal O}(\sqrt{N_{\rm det}})$ scaling in this case, we have
measured the CPU time needed to realize one Monte Carlo step. 
All the determinants are computed with the cubic algorithm (no Sherman-Morrison
updates). However, results obtained with SM updates show exactly the same behavior. 
The timings are given in table~\ref{tab:cl_dz:1}, together with the
number of $\uparrow$ and $\downarrow$ unique determinants, the number of occupied
molecular orbitals, and the amount of RAM needed per CPU core. The
computational cost compared to a single-determinant calculation is given in
figure~\ref{fig:scaling}.

\begin{table}
\begin{tabular}{|c|c|c|c|c|c|}
\hline
 $N_{\rm det}$ & $N_{\rm det}^\uparrow$ & $N_{\rm det}^\downarrow$ & $N_{\rm inv}$ & $N_{\rm subst}$ & CPU time(ms)  \\ 
\hline
       1       &          1           &          1          &     2.0     &    0    &    0.0179     \\ 
       10      &          7           &          7          &     2.0     &    29   &    0.0237     \\ 
      100      &          40          &          32         &     2.1     &   130   &    0.0428     \\ 
     1 000     &         250          &         186         &     2.6     &   925   &    0.1643     \\ 
     10 000    &        1 143         &         748         &     5.1     &  4 283  &    0.6808     \\ 
    100 000    &        5 441         &        3 756        &     17.7    & 19 049  &    3.2794     \\ 
   1 000 000   &       21 068         &       14 516        &     54.7    & 63 325  &   12.8688     \\ 
\hline
\end{tabular}
\caption{Number of determinants ($N_{\rm det}$, $N_{\rm det}^\uparrow$ and
$N_{\rm det}^\downarrow$), average number of matrix inversions in ${\cal O}(N^3)$
($N_{\rm
inv}$), average number of Sherman-Morrison column substitutions ($N_{\rm subst}$)
and CPU time per core per Monte Carlo step.}
\label{tab:cl_dz:2}
\end{table}

From the data of table \ref{tab:cl_dz:1} one can observe a CPU time
scaling almost perfectly linearly with $N_{\rm det}^\uparrow+N_{\rm det}^\downarrow$. 
Indeed, fitting the data with a law of the form
$c N_{det}^\gamma$, the value of $\gamma$ obtained is 1.02.
Now, regarding the scaling obtained with respect to the total
number of determinant products, we found in all cases (tables \ref{tab:cl_dz:1},\ref{tab:cl_dz:2} 
and for the two spin sectors) an exponent $\gamma$ around 0.6. This value is 
slightly higher than the expected $\gamma=0.5$ due to the sparsity
of the ${\bf C}$ matrix: among all possible determinant products, many have
a zero coefficient and those are not counted in $N_{\rm det}$.

\subsection{Speedup due to Sherman-Morrison updates}

Now, including the Sherman-Morrison updates, a speedup of $6-7\times$  is
obtained. According to the documentation of the MKL library, the full matrix
inversion ({\tt dgetrf} followed by {\tt dgetri}) uses approximately $2n^3$
floating-point (FP) operations for an $n\times n$ matrix, whereas one
Sherman-Morrison column substitution in our implementation uses $5 n^2 + 2 n +
3$ FP operations. From the data of table~\ref{tab:cl_dz:2}, the average number
of Sherman-Morrison updates per determinant ranges between 1.7 and 2.5. From
these results, one can conclude that our implementation of column substitutions
has an efficiency higher than the efficiency of the matrix inversion using the
MKL library for such small matrices: for an $9\times 9$ matrix, two column
substitution involve 1.71 times less FP operations than the full matrix
inversion, which is four times less than the speedup we measure.

\begin{table}
\begin{tabular}{|c|c|c|c|}
\hline
 Excitation & $D^\uparrow D^\downarrow$ &  $D^\uparrow$ &  $D^\downarrow$  \\ 
\hline
         0         &           1            &           1            &           1            \\ 
         1         &           38           &           90           &           88           \\ 
         2         &         2 177          &          1603          &          1520          \\ 
         3         &         43 729         &          7811          &          6507          \\ 
         4         &        308 045         &          8581          &          5071          \\ 
         5         &        351 182         &          2090          &          579           \\ 
         6         &        291 481         &           77           &           0            \\ 
         7         &         3 067          &           0            &           0            \\ 
         8         &          280           &           0            &           0            \\ 
\hline
\end{tabular}
\caption{In the 1~000~000-determinant wave function, the number of
determinants resulting from excitation operators of degree 0 to 8
applied on the Hartree-Fock reference ($D^\uparrow_1D^\downarrow_1$,
$D^\uparrow_1$ or $D^\downarrow_1$).}
\label{tab:cl_dz:exc}
\end{table}

The average number of substitutions is lower than what one would have obtained
with a fixed reference determinant. For instance, if the Hartree-Fock
$\uparrow$ and $\downarrow$ determinants had been taken as a fixed reference and
assuming all substitutions were successful (with a determinant ratio greater
than $10^{-3}$ in absolute value), the average number of Sherman-Morrison
substitutions would have been equal to 3.46 according to the data of
table~\ref{tab:cl_dz:exc} where we have measured an average of 1.78 for
the same wave function with our implementation.

The average number of matrix inversions per step is at least two since one
determinant has to be computed for each spin. Then, the probability of using
the ${\cal O}(N^3)$ algorithm instead of the Sherman-Morrison updates to
reduce the propagation of numerical errors stays very low below 0.17~\%.
We have checked that for a given set of electron coordinates the local energies
computed with and without the Sherman-Morrison updates differ by no more
than $2\, 10^{-5}$ atomic units on all the wave functions.  These data confirm
that the numerical stability of the Sherman-Morrison updates can be controlled
without affecting significantly the computational time.

\subsection{Use of the improved truncation scheme}

\begin{table}
\begin{tabular}{|l|r|r|r|c|}
\hline
 Threshold $\epsilon$ & $N_{\rm det}$ & $N_{\rm det}^\uparrow$ & $N_{\rm det}^\downarrow$ & CPU time(ms)  \\ 
$|c_k| > \epsilon$  &               &                      &                     &               \\ 
\hline
       0.96       &       1       &          1           &          1          &   0.02450     \\ 
      0.0404      &       10      &          8           &          8          &   0.03080     \\ 
      0.0103      &      100      &          53          &          35         &   0.05888     \\ 
 $2.02\,10^{-2}$  &      110      &          54          &          37         &   0.06004     \\ 
 $10^{-2}$        &      1000     &         254          &         168         &    0.1745     \\ 
 $2.53\,10^{-3}$  &      2003     &         496          &         335         &    0.3364     \\ 
 $10^{-3}$        &     10000     &         1700         &         994         &    0.9732     \\ 
    $10^{-4}$     &     30198     &         3668         &         1853        &    1.920      \\ 
 $3.55\,10^{-5}$  &     100000    &         9256         &         4524        &    4.912      \\ 
    $10^{-5}$     &     348718    &        24758         &        12511        &    14.20      \\ 
    $10^{-6}$     &     993811    &        52291         &        26775        &    31.11      \\ 
       0.0        &    1000000    &        52433         &        26833        &    31.92      \\ 
\hline
\hline
 Threshold $\epsilon_\uparrow=\epsilon_\downarrow=\epsilon$ & $N_{\rm det}$ & $N_{\rm det}^\uparrow$ & $N_{\rm det}^\downarrow$ & CPU time(ms)  \\ 
${\cal N}_i^\uparrow > \epsilon \; ;\; {\cal N}_j^\downarrow > \epsilon$&       &    &    &               \\ 
\hline
  $10^{-2}$  &       1       &          1           &          1          &   0.02450     \\ 
 $5.10^{-3}$ &       3       &          3           &          2          &   0.02688     \\ 
  $10^{-3}$  &       86      &          21          &          17         &   0.03899     \\ 
 $5.10^{-4}$ &      214      &          29          &          28         &   0.04636     \\ 
  $10^{-4}$  &      1361     &          93          &          74         &   0.08808     \\ 
 $5.10^{-5}$ &      2424     &         120          &          89         &    0.1054     \\ 
  $10^{-5}$  &      9485     &         234          &         166         &    0.1855     \\ 
  $10^{-6}$  &     54016     &         772          &         523         &    0.5960     \\ 
  $10^{-7}$  &     207995    &         2279         &         1389        &    1.740      \\ 
  $10^{-8}$  &     459069    &         5797         &         3291        &    4.196      \\ 
  $10^{-9}$  &     748835    &        14456         &         8054        &    9.724      \\ 
  $10^{-10}$ &     926299    &        30320         &        16571        &    19.18      \\ 
      0      &    1000000    &        52433         &        26833        &    31.92      \\ 
\hline
\end{tabular}
\caption{Number of determinants, number of spin-specific determinants and computational
cost as a function of the truncation threshold with two different truncation approaches. The cc-pVTZ 
basis sets is used.}
\label{tab:truncation}
\end{table}

We have generated a wave function for the chlorine atom using the cc-pVTZ basis set
with one million determinants. The determinants are generated with the CIPSI
algorithm in the FCI space with 2 frozen electrons in the $1s$ orbital
(2 MOs always doubly occupied, and 15 active electrons in 37 MOs).
This wavefunction has been truncated using a standard truncation scheme based on the absolute
value of the CI coefficients 
(products of determinants in Eq.(\ref{eq1}) with $|c_k| \le \epsilon $ are removed)
and using the contribution of spin-specific determinants
to the norm of the wavefunction as proposed in the preceding section. 
The wave functions obtained after truncation as well as the computational time in 
milliseconds per Monte Carlo step are detailed in table \ref{tab:truncation}.

As it should be the timings in the case of one single or all determinants are identical in both cases. 
Choosing $\epsilon$ large enough, only one determinant is kept. By decreasing the threshold 
the number of determinants increases, but with a marked difference between the two truncation schemes. 
For a given total number of determinants $N_{\rm det}$, the proposed scheme contains much less spin-specific 
determinants than in the standard case. For example, for $N_{\rm det} \sim 100$ there are about two times 
less $\sigma$-determinants. For $N_{\rm det} \sim 2~000$ the factor is about 5 and close to 
$N_{\rm det} \sim 10~000$ a factor 7 is observed. Furthermore, the gains in CPU times evolve with 
the same factor since the computational time is proportional to the number of spin-specific
determinants (not $N_{\rm det}$).

\begin{figure}
\begin{center}
\includegraphics[width=\columnwidth]{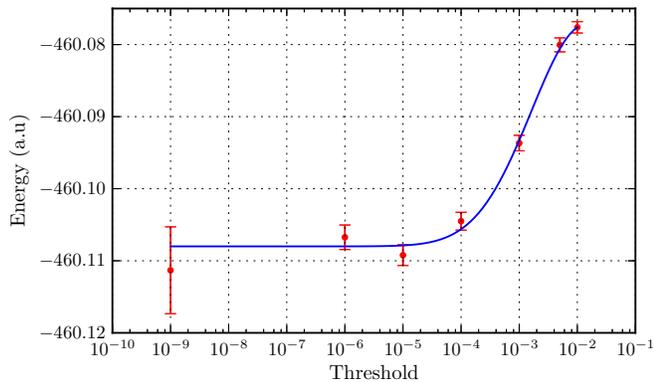}
\end{center}
  \caption{     \label{fig:energy}
Total FN-DMC energy of the chlorine atom with truncated near-Full-CI/cc-pVTZ
wave functions. Spin-specific determinants with a contribution to the norm less
than the threshold are removed.
  }
\end{figure}

\begin{table}
  \begin{tabular}{|c|c|c|}
\hline
    Threshold     & Energy (a.u.)  & CPU time (hours) \\
    \hline
    $10^{-2}$     &   -460.0776(08)   & 15 820 \\
 $5\,10^{-3}$     &   -460.0800(10)   & 14 521 \\
    $10^{-3}$     &   -460.0937(11)   & 14 522 \\
    $10^{-4}$     &   -460.1045(12)   & 19 109 \\
    $10^{-5}$     &   -460.1092(14)   & 22 344 \\
    $10^{-6}$     &   -460.1067(17)   & 47 679 \\
    $10^{-9}$     &   -460.1113(60)   & 48 530 \\
    \hline
  \end{tabular}
\caption{All-electron Fixed-Node DMC energies.  The threshold is applied to the contribution of the spin-specific determinants to the norm of the wavefunction.}
\label{tab:energy}
\end{table}

On figure~\ref{fig:energy}, we
can see that the FN-DMC energy is converged within the error bars with a
threshold of $10^{-6}$ on the contributions ${\cal N}_i^\uparrow$ and ${\cal
N}_j^\downarrow$ to the norm of the wave function.
Table~\ref{tab:energy} gives the energies of the truncated wave functions, and
the CPU time needed to run the calculations.  

\begin{figure}
\begin{center}
\includegraphics[width=\columnwidth]{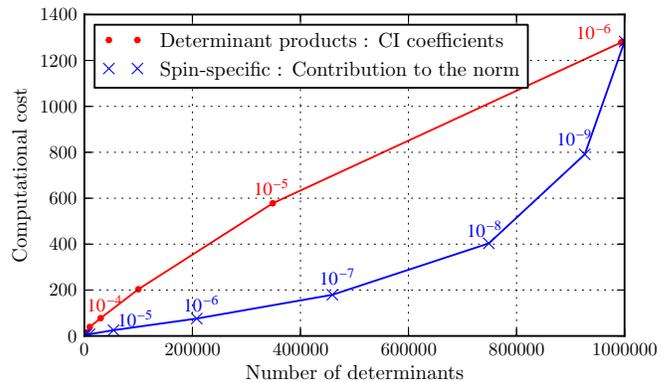}
\end{center}
  \caption{
  \label{fig:truncation}
    Computational cost with respect to the number of determinants, normalized to
    the cost of a single-determinant calculation. The truncation is applied to
    the absolute value of the CI coefficients or to the contribution to the norm of
    the spin-specific determinants. The value of the truncation threshold is given
    on the figure next to the corresponding points.
  }
\end{figure}

Figure~\ref{fig:truncation} compares the computational cost obtained with the two
truncation approaches. For a fixed computational cost, our truncation strategy
keeps a much larger number of determinant products than the standard truncation
scheme. Removing spin-specific determinants which contribute to less than
$10^{-5}$ of the norm can make the calculation of a wave function with 9~485
determinants cost only 7.6 more than a single determinant calculation.
Note that with such a number the FN-DMC energy is expected to be converged as a function of 
the number of Slater determinants (see, Fig.\ref{fig:energy}).

\section{Some remarks about comparisons with other methods}
\label{sec5}
Comparing the {\it practical} performance of various methods is not easy. 
A first important aspect to consider is the formal scaling of the computational cost
(both in terms of CPU and memory) as a function of the critical parameters:
Number of electrons, molecular orbitals, and determinants. However, 
some caution is required since the asymptotic regime where 
such scaling laws are valid is not necessarily reached for the range of values considered. 
Here, we are in such a situation since a square-root law for the computational cost 
as a function of the number of determinants is approximately observed up to about one million 
of determinants, despite the fact that the theoretical dependence is linear. 

A second important aspect -- which is in general underestimated -- 
is the importance of optimally exploiting 
the high-performance capabilities of present-day processors. 
Such an aspect must be taken into consideration not only 
when implementing a given algorithm but also, and 
much more importantly, at the moment of deciding what type of 
algorithm should be used to achieve the desired calculations (algorithm design step).
While describing our algorithm we have mentioned several 
important features. The use of vector fused-multiply add (FMA) 
instructions (that is, the calculation of \texttt{a=a+b*c} in one CPU cycle) for the innermost loops is extremely 
efficient and should be searched for. Using such instructions (present in general-purpose processors),
up to eight FMA per CPU cycle can be performed. While computing loops, overheads 
are also very costly and should be reduced/eliminated. 
By taking care separately of the various parts of the loop
(peeling loop, scalar loop, vector loop, and tail loop) through size-specific and/or 
hard-coded subroutines, a level of 100\% vectorized loops can be reached. 
Another crucial point is to properly manage the data flow arriving to the processing unit. 
As known, to be able to move data from the memory to the CPU with a sufficiently high data transfer to 
keep the CPU busy is a major concern of modern calculations. 
Then, it is not only important to make maximum use of the low-latency cache memories to 
store intermediate data but also to maximize prefetching allowing
the processor to anticipate the use of the right data and instructions in advance. 
To enhance prefetching the algorithm should allow the predictability 
of the data arrival in the CPU (that is, avoid random access as much as possible). 
All these various practical aspects 
are far from being anecdotal since they may allow orders of magnitude in computational savings.
We emphasize that in this work we have chosen to make use of Sherman-Morrison (SM) updates,
despite the fact that it is not the best approach in terms of formal scaling (for example, 
the Table method discussed below has a better scaling). However, the massive calculations
of scalar products at the heart of repeated uses of SM updates are so ideally adapted to 
the features of present-day processors just described above, that very high performances can 
be obtained.

To give a quantitative illustration of such ideas we present now some comparisons 
between the timings obtained with our algorithm and those obtained with 
the Table method of Clark {\it et al.},\cite{Clark_2011} one of the most efficient approach proposed so far. 
It is clear that making {\it fair} comparisons between algorithms implemented 
within different contexts by different people 
is particularly difficult. Accordingly, the timings given below must be taken 
with lot of caution and should just be understood as an illustration of the main issues.
Ultimately, it is preferable to compare the actual timings 
obtained for a given application, with a given code, and a given processor. In this spirit, 
we present in tables \ref{tab:cl_dz:1} and \ref{tab:cl_dz:2} our timings (in ms)
for an elementary Monte Carlo step (all electrons moved once) in the case of the Cl atom.

We have coded the Table method in the QMC=Chem code. In brief,
the approach consists in computing the $N_{det}^\sigma$ determinants, Det${\bf S}^\sigma_k$,
and their derivatives, from the evaluation of the series of ratios 
Det${\bf S}^\sigma_k$/Det${\bf S}^\sigma_0$,
where ${\bf S}^\sigma_0$ is the reference Slater matrix.
Denoting $s$ the number of particle-hole excitations connecting ${\bf S}^\sigma_0$ and 
${\bf S}^\sigma_k$ the ratio of determinants can be expressed as 
the determinant of a small $s$x$s$ matrix 
whose matrix elements are taken from a larger table of size $N_\sigma$x$M$ computed in a 
preliminary step
($M$= number of virtual orbitals used in the expansion).
The main computational costs are the reading of the $s^2$ elements 
in the pre-computed table and the computation of the determinants of size $s$ with a $s^3$ cost, 
the two steps being performed for each elementary determinant. 
In theory, the algorithm is attractive since it avoids the repeated computation 
of SM updates whose cost increases as the square 
of the number of $\sigma$-electrons. However, in practice this advantage can be 
counterbalanced by the cost of making expensive (partially) random 
access to the table. It is particularly true in the case where large numbers
of electrons and/or basis set are used, a situation where 
the entire table cannot be stored in the lowest-level cache.

To quantify such aspects, we present now some measurements of the cost of the main steps 
of both algorithms expressed in number of CPU cycles. The task considered is the calculation 
of a wavefunction consisting of a total of $N_{\rm det}$= 926299 determinants and 
involving  30320 different $\uparrow$-determinants
and 16571 $\downarrow$-determinants
(this is the wavefunction corresponding to the 
threshold $\epsilon_\uparrow=\epsilon_\downarrow=10^{-10}$ in Table \ref{tab:truncation}). 
Note that time measurements are accurate with a precision 
of about $\pm 20$ cycles. For the processor used here, a CPU cycle time is equal to $t_{{\rm CPU}}=$0.28~ns. 

Using our optimized SM algorithm, the time spent to the computation 
of the initial reference inverse matrix and determinant 
for the 9 $\uparrow$-electrons and 8 $\downarrow$ electrons is measured to be 8368 and 7730 
cycles, respectively. 
In average (over the all set of different determinants), the cost of updating
the inverse Slater matrix and computing the determinant is found to be about 
544 and 359 cycles, respectively. The average number of substitutions (and, thus, number of elementary 
SM step corresponding to one-column substitution) being about $N_{subst} \sim 1.8$, 
a rough estimate of the total cost is then
$$
T_{\rm SM} = [8368+7730+ (544 N^{\uparrow}_{\rm det} + 359 N^{\downarrow}_{\rm det} ) N_{\rm subst}] t_{\rm CPU}
\sim  11.3 {\rm ms}
$$
in good qualitative agreement with the total timing of 19.2 ms obtained by direct measurement 
and reported in Table 5. 

In the case of the Table method the initial step consisting in evaluating 
the inverse of ${\bf S}^\sigma_0$ and its determinant on one hand and 
constructing the table on the other hand, have been measured to take 16140 and 14574 cycles, 
respectively (sum of $\uparrow$ and $\downarrow$ contributions).
For each ratio to evaluate, reading the table
and calculating the determinant of the $s\times s$ matrix using the LAPACK routine 
\texttt{dgetrf}, are found to take in average 2430 and 2187 cycles, for each spin respectively. 
An estimate of the cost is thus 
$$
T_{{\rm Tab}}= [16140 + 14574 + (2430 N^{\uparrow}_{\rm det} + 2187 N^{\downarrow}_{\rm det})] t_{{\rm CPU}}
\sim 30.8 {\rm ms}
$$
In this case it is seen that the optimized SM algorithm is approximately three times faster than the 
Table method. Although this schematic comparison should be taken with lot of caution, it nevertheless
illustrates that our optimized SM algorithm is a competitive algorithm. 
It should also be noted that in the present case, 
the table of the Table method is sufficiently small (518 matrix elements) to be entirely stored in the 
low-latency L1 cache. For larger numbers of electrons and basis sets, it will be 
no longer true and important additional times should be lost because of the numerous (partially) 
random access to higher-level memories.


\section{Summary}
\label{conclu}

The objective of this work was to present in detail our algorithm for computing 
very efficiently large multideterminant expansions. As illustrated here 
for the chlorine atom and elsewhere in other applications,\cite{Giner_2015,Scemama_2014,Giner_2013,Caffarel_2014} this algorithm allows to realize converged FN-DMC simulations 
using a number of determinants superior to what has been presented so far 
in the literature. For the chlorine atom presented here, FN-DMC calculations
using about 750~000 determinants with a computational increase of only $\sim$ 400 compared
to a single-determinant calculation have been shown feasible.
Several aspects make this algorithm particularly efficient. 
They include not only algorithmic improvements but also very practical considerations about the 
way the calculations are implemented on present-day processors. We strongly emphasize that 
this last aspect is by no way anecdotal and must absolutely be taken into account when 
an efficient algorithm has to be devised and implemented. 
Our experience shows that orders of magnitude in efficiency 
can be gained by taking this aspect into consideration. 
Here, the choice of using SM updates instead of a more elegant 
scheme (such as, for example, the Table method of Clark {\it et al.}\cite{Clark_2011} that 
has a better formal scaling) has been driven 
by the fact that massive computations of scalar products are ideally suited to 
modern processors and can be performed extremely efficiently.

As just said, we calculate the determinants and their derivatives 
using the Sherman-Morrison formula for updating the inverse Slater matrices, 
as proposed in a number of previous works. 
In contrast with other implementations, we have found more efficient
not to compare the Slater matrix to a common reference (typically, the Hartree-Fock determinant)
but instead to perform the Sherman-Morrison updates with respect to the
previously computed determinant $D_{i-1}^\sigma$.
To reduce the prefactor associated with this step
we have sorted the list of determinants with a suitably chosen order
so that with high probability
successive determinants in the list differ only by one- or two-column substitution, thus
decreasing the average number of substitution performed.

In this work, we have emphasized that multideterminant expansions contain in general
a large number of identical spin-specific determinants [for
typical configuration interaction-type wavefunctions
the number of unique spin-specific determinants $N_{\rm det}^\sigma$ ($\sigma=\uparrow,\downarrow$),
having a non-negligible weight in the wavefunction is of order $O(\sqrt{N_{\rm det}})$].
To have the full benefit of this remark, that is, to get in practice a square-root law 
over a wide range of numbers of determinants, it is essential to be able to keep negligible 
the contributions whose cost scales with the total number of determinants. 
As described in the two sections 
devoted to the computation of the intermediate vectors and 
the gradients and Laplacian, the computationally intensive parts of 
such contributions can be mainly restricted to the calculation of two matrix-vector products, 
performed only once for the wavefunction and the $6 N_{el}$ derivatives. 
A number of technical details related to the 
way such a calculation should be efficiently implemented on a modern process have been given. 

Finally, by taking advantage of the bilinear form 
for the multideterminant expansion, Eq.(\ref{eq:main}), 
a new truncation scheme has been proposed. Instead of truncating the expansion 
according to the magnitude of the coefficients of the expansion as usual, we propose 
to remove spin-specific determinants instead according to their total contribution to the norm 
of the expansion. In this way, more determinants can be handled for a price corresponding to shorter 
expansions.

\subsection*{Acknowledgments}

AS and MC thank the Agence Nationale pour la Recherche (ANR) for support 
through Grant No ANR 2011 BS08 004 01.
This work has been possible thanks to the computational support of CALMIP (Toulouse), and GENCI projects x2015067347 and x2015081738.

\bibliographystyle{unsrt}
\bibliography{arxiv}

\begin{thebibliography}{10}

\bibitem{Giner_2015}
Emmanuel Giner, Anthony Scemama, and Michel Caffarel.
\newblock {Fixed-node diffusion Monte Carlo potential energy curve of the
  fluorine molecule F2 using selected configuration interaction trial
  wavefunctions}.
\newblock {\em J. Chem. Phys.}, 142(4):044115, jan 2015.

\bibitem{Scemama_2014}
Anthony Scemama, Thomas Applencourt, Emmanuel Giner, and Michel Caffarel.
\newblock {Accurate nonrelativistic ground-state energies of 3d transition
  metal atoms}.
\newblock {\em J. Chem. Phys.}, 141(24):244110, dec 2014.

\bibitem{Giner_2013}
Emmanuel Giner, Anthony Scemama, and Michel Caffarel.
\newblock {Using perturbatively selected configuration interaction in quantum
  Monte Carlo calculations}.
\newblock {\em Canadian Journal of Chemistry}, 91(9):879--885, sep 2013.

\bibitem{Caffarel_2014}
Michel Caffarel, Emmanuel Giner, Anthony Scemama, and Alejandro
  Ram{\'{\i}}rez-Sol{\'{\i}}s.
\newblock {Spin Density Distribution in Open-Shell Transition Metal Systems: A
  Comparative Post-Hartree{\textendash}Fock Density Functional Theory, and
  Quantum Monte Carlo Study of the {CuCl} 2 Molecule}.
\newblock {\em J. Chem. Theory Comput.}, 10(12):5286--5296, dec 2014.

\bibitem{Nukala_2009}
Phani K. V.~V. Nukala and P.~R.~C. Kent.
\newblock A fast and efficient algorithm for slater determinant updates in
  quantum monte carlo simulations.
\newblock {\em The Journal of Chemical Physics}, 130(20):204105, 2009.

\bibitem{Clark_2011}
Bryan~K. Clark, Miguel~A. Morales, Jeremy McMinis, Jeongnim Kim, and Gustavo~E.
  Scuseria.
\newblock {Computing the energy of a water molecule using multideterminants: A
  simple efficient algorithm}.
\newblock {\em J. Chem. Phys.}, 135(24):244105, 2011.

\bibitem{Weerasinghe_2014}
Gihan~L. Weerasinghe, Pablo~L{\'{o}}pez R{\'{\i}}os, and Richard~J. Needs.
\newblock {Compression algorithm for multideterminant wave functions}.
\newblock {\em Physical Review E}, 89(2), feb 2014.

\bibitem{note1}
The best gain in computational savings given in \cite{Weerasinghe_2014} is
  26.57 ($N_s/N_d$ for B in Table I). However, it includes a factor of about
  2.5 associated with the decompression of the wavefunction written in terms of
  CSF's and the regrouping of identical determinants. For the sake of
  comparison, this latter factor should not be taken into account since the
  starting multideterminant wavefunction used here is supposed to be expanded
  over a set of {\it different} determinants.

\bibitem{Scemama_2013}
Anthony Scemama, Michel Caffarel, Emmanuel Oseret, and William Jalby.
\newblock {{QMC}=Chem: A Quantum Monte Carlo Program for Large-Scale
  Simulations in Chemistry at the Petascale Level and beyond}.
\newblock In {\em High Performance Computing for Computational Science -
  {VECPAR} 2012}, pages 118--127. Springer Science Business Media, 2013.

\bibitem{notesqrt}
This general result can be illustrated in the particular case of the chlorine
  atom considered here. Using our perturbatively selected CI wavefunction
  including up to one-million of determinants (cc-pVDZ basis set), the number
  of unique spin-specific determinants $N^\sigma_{det}$ (see, table
  \ref{tab:cl_dz:1}) is observed to scale roughly as the square root of the
  total number of determinants for both spins. Indeed, a least-square fit of
  the data using a law of the form $c N_{det}^\gamma$ leads to about $\gamma=
  0.6$ for both spin sectors. Let us now consider a CISD calculation in the
  same basis set. Using a value of $10^{-9}$ as threshold for the coefficients,
  the total contribution corresponding to the HF, single-, and
  double-excitations are found to be about 0.936, 0.002, and 0.062,
  respectively. Among all double excitations present in the expansion, those
  involving two electrons of opposite spins contribute for a total of about
  0.050, while spin-like excitations contribute only for 0.008 and 0.004 for
  the $\uparrow$ and $\downarrow$ sector, respectively.

\bibitem{Clay_2015}
Raymond~C. Clay and Miguel~A. Morales.
\newblock {Influence of single particle orbital sets and configuration
  selection on multideterminant wavefunctions in quantum Monte Carlo}.
\newblock {\em J. Chem. Phys.}, 142(23):234103, jun 2015.

\bibitem{Scemama_2013_2}
Anthony Scemama, Michel Caffarel, Emmanuel Oseret, and William Jalby.
\newblock {Quantum Monte Carlo for large chemical systems: Implementing
  efficient strategies for petascale platforms and beyond}.
\newblock {\em J. Comput. Chem.}, 34(11):938--951, jan 2013.

\bibitem{slater_condon}
Anthony Scemama and Emmanuel Giner.
\newblock {An efficient implementation of Slater-Condon rules}.
\newblock {\em ArXiv e-prints}, nov 2013.

\bibitem{Andersson_1995}
Arne Andersson, Torben Hagerup, Stefan Nilsson, and Rajeev Raman.
\newblock {Sorting in linear time?}
\newblock In {\em Proceedings of the twenty-seventh annual {ACM} symposium on
  Theory of computing - {STOC} 95}. {ACM} Press, 1995.

\bibitem{lapack}
E.~Anderson, Z.~Bai, C.~Bischof, S.~Blackford, J.~Demmel, J.~Dongarra,
  J.~Du~Croz, A.~Greenbaum, S.~Hammarling, A.~McKenney, and D.~Sorensen.
\newblock {\em {{LAPACK} Users' Guide}}.
\newblock Society for Industrial and Applied Mathematics, Philadelphia, PA,
  third edition, 1999.

\bibitem{maqao}
L.~Djoudi, D.~Barthou, P.~Carribault, C.~Lemuet, J.-T. Acquaviva, and W.~Jalby.
\newblock {MAQAO: Modular assembler quality Analyzer and Optimizer for Itanium
  2}.
\newblock In {\em {Workshop on EPIC Architectures and Compiler Technology San
  Jose, California, United-States}}, Mar 2005.

\bibitem{Woon_1993}
David~E. Woon and Thom~H. Dunning.
\newblock {Gaussian basis sets for use in correlated molecular calculations.
  {III}. The atoms aluminum through argon}.
\newblock {\em J. Chem. Phys.}, 98(2):1358, 1993.

\bibitem{quantum_package}
A.~Scemama, E.~Giner, T.~Applencourt, G.~David, and M.~Caffarel.
\newblock Quantum package v0.6, September 2015.
\newblock doi:10.5281/zenodo.30624.

\end{thebibliography}

\end{document}